\documentstyle[12pt]{article}
\def\del{\partial}
  \makeatletter

        \@addtoreset{equation}{section}
        \def\section{\@startsection {section}{1}{\z@}{3.5ex plus -1ex minus
        -.2ex}{2.3ex plus .2ex}{\large\bf}}
  \makeatother
\begin{document}
\setlength{\baselineskip}{22pt}
\rightline{YAMAGATA-HEP-93-14}
\rightline{August 1993}
\vspace{1.5truecm}
\centerline{\Large\bf An Extension of Type I Gaugeon Formalism}
\centerline{\Large\bf for Quantum Electrodynamics}
\vspace{2truecm}
\centerline{Ryusuke ENDO}
\bigskip
\centerline{\it Department of Physics, Yamagata University, Yamagata 990,
Japan}
\bigskip
\vspace{2truecm}
\centerline{\bf Abstract}
\bigskip
By introducing two kinds of gaugeon fields, we extend Yokoyama's Type I
gaugeon formalism for quantum electrodynamics. The theory admits a q-number
gauge transformation by which we can shift the gauge parameter into
arbitrary numerical value; whereas in the original theory we cannot change
the sign of the parameter. The relation to the Type II theory is also
discussed.

\thispagestyle{empty}
\newpage
\section{Introduction}

In the standard formalism of canonically quantized gauge theories
\cite{N,KO}
we cannot consider the gauge transformation freely.
There exists no gauge freedom in the quantum theory, since the quantum
theory is defined only after the gauge fixing. Namely, a Hilbert space
defined in a particular gauge is quite different from those in other gauges.
Thus, if we want to realize the quantum gauge freedom, we need a wider
Hilbert space.

Yokoyama's gaugeon formalism \cite{OEM}--\cite{YM2} provides a wider
framework in which we can consider the quantum gauge transformation
among a family of Lorentz covariant linear gauges. In this formalism
a set of extra fields, so called gaugeon fields, is introduced as the quantum
gauge freedom. This theory was first proposed for the quantum
electrodynamics \cite{OEM}--\cite{txt}
to resolve the problem of gauge parameter renormalization \cite{HY}.
It was also applied later to the Yang-Mills theory \cite{YM1,YM2}.
Thanks to the quantum gauge freedom of this formalism, the gauge parameter
independence of the physical $S$-matrix becomes manifest \cite{Smatrix}.
It has also been shown, with the help of certain conjecture,
that the wave-function renormalization constant is gauge independent
in this formalism \cite{Zfactor}.

There are two types of the gaugeon theory, which Yokoyama and Kubo \cite{YK}
called Type I and Type II theory.
The Lagrangian of the Type I gaugeon formalism for electromagnetic field
$A_\mu$ coupled to charged field $\psi$ is given by
\begin{equation}
  {\cal L}_{\rm I} = {\cal L}_{\rm inv}
                   + \partial_\mu B A^\mu
                   + {\varepsilon \over 2} \left( Y_* + \alpha B \right)^2
                   + \partial_\mu Y_* \partial^\mu Y,
                                              \\   \label{type1}
\end{equation}
and the Lagrangian of Type II theory is
\begin{equation}
   {\cal L}_{\rm II}= {\cal L}_{\rm inv}
                    + \partial_\mu B A^\mu + {\alpha \over 2}B^2
                    + {1\over2} Y_* B
                    + \partial_\mu Y_* \partial^\mu Y,
                                             \label{type2}
\end{equation}
with ${\cal L}_{\rm inv}$ being the gauge invariant Lagrangian
\begin{equation}
     {\cal L}_{\rm inv} = - {1 \over 4}F_{\mu\nu}F^{\mu\nu}
                          + {\cal L}_{\rm matt}(\psi, A_\mu),
                                           \label{Linv}
\end{equation}
where $F_{\mu\nu}=\partial_\mu A_\nu-\partial_\nu A_\mu$, $B$ is the
$B$-field of Nakanishi-Lautrup \cite{N},
$Y$ and $Y_*$ are the gaugeon field and its associated field,%
\footnote{
               We use field notation different from Yokoyama's;
               the fields $B$, $Y$ and $Y_*$ here are denoted by $B_1$,
               $B$ and $B_2$, respectively, in the original paper \cite{OEM}.
                           }
$\alpha$ is the gauge parameter in this formalism and ${\cal L}_{\rm matt}$
is the Lagrangian for the matter field $\psi$ minimally coupled to $A_\mu$.
The parameter $\varepsilon$ in Type I Lagrangian (\ref{type1}) is a sign
factor ($\varepsilon=\pm1$). Namely, Type I theory can be classified further
into two types,
which we denote in this paper by Type I$_+$ and Type I$_-$
corresponding to the sign $\varepsilon$.
The gauge parameter of the standard formalism, which we denote by $a$ in the
present paper, can be identified with $a=\varepsilon\alpha^2$ for Type I theory
and with $a=\alpha$ for Type II theory.
For example, these Lagrangians yield the photon propagator as
\begin{equation}
           {{g_{\mu\nu}}\over k^2}
                + (a - 1){k_\mu k_\nu \over (k^2)^2}
                                  \label{propagator}
\end{equation}
with
\begin{eqnarray}
           &a=\varepsilon \alpha^2 &\quad \mbox{ for Type I}, \label{aI} \\
           &a=\alpha ~             &\quad \mbox{ for Type II}.\label{aII}
\end{eqnarray}
In particular, $\alpha=0$ corresponds to Landau gauge ($a=0$) and $\alpha=1$
(with $\varepsilon=+1$ for Type I) gives Feynman gauge ($a=1$).

Both Lagrangians admit the q-number gauge transformation. Under the field
transformation
\begin{eqnarray}
   & &\hat A_\mu = A_\mu + \tau \partial _\mu Y, \nonumber \\
   & &\hat \psi = e^{i\tau eY} \psi,             \nonumber \\
   & &\hat Y_* = Y_* - \tau B,                   \nonumber \\
   & &\hat B = B, \quad \hat Y = Y,             
                                                  \label{qgt1}
\end{eqnarray}
with $\tau$ being a parameter, the Lagrangians is {\it form invariant}, that
is, they transform as
\begin{equation}
             {\cal L}_{\rm I,II}(\phi^A; \alpha)
                =  {\cal L}_{\rm I,II}(\hat \phi^A; \hat \alpha),
                                \label{forminvariance}
\end{equation}
where 
${\phi^A}$ stands for any of the fields and $\hat\alpha$ is defined by
\begin{equation}
          \hat \alpha = \alpha + \tau .   \label{hat_alpha}
\end{equation}
The form invariance (\ref{forminvariance}) means that $\phi^A$ and
$\hat \phi^A$ satisfy the same field equation except for the parameter
$\alpha$ which should be replaced by $\hat \alpha$ for the
$\hat \phi^A$ field equation. Thus, we can shift the gauge parameter by
the q-number gauge transformation. The
sign factor $\varepsilon$ in Type I theory, however, cannot be changed by any
transformation.

We should ensure that the extra gaugeon modes do not contribute to the
physical processes.
In fact, the gaugeon fields exhibit dipole character and thus
yield negative normed states that would lead to the negative
probability \cite{OEM}. To remove these unphysical gaugeon modes Yokoyama
imposed a Gupta-Bleuler type subsidiary condition \cite{OEM}, which is not
applicable if interaction exists for gaugeon fields. Yokoyama's subsidiary
condition has been improved by introducing BRST symmetry for gaugeon fields
in Ref.\cite{Izawa} for Type II theory and in Ref.\cite{KSE} for Type I
theory. In these BRST symmetric gaugeon formalisms, unphysical gaugeon modes
as well as unphysical photons are removed by the single Kugo-Ojima type
condition which is applicable even if interaction exists.

Comparing both types of the formalism with each other, we observe the
following aspects:
\begin{enumerate}
 \item
  Although the shift of $\alpha$ (\ref{hat_alpha}) is common for both type
  theories, the standard gauge parameter $a$ appearing in the propagator
  shifts differently:
  \begin{eqnarray}
   &\hat a = \varepsilon(\alpha + \tau)^2, \quad & \mbox{ for Type I}
                                                 \nonumber \\
   &\hat a = \alpha + \tau = a + \tau , \quad & \mbox{ for Type II}.
                                                 \nonumber
  \end{eqnarray}
  We can shift the parameter $a$ into arbitrary value in the Type II theory.
  In the Type I theory, however, we cannot change the sign of $a$.
 \item
  If we put $\alpha=0$ in the Type I Lagrangian (\ref{type1}), the gaugeon
  sector decouples from the rest. Then the remaining sector has the same form
  with the Lagrangian of the standard formalism in Landau gauge. Thus, the
  equivalence of the theory to the standard formalism is manifest in Landau
  gauge.
  This situation does not occur in the Type II theory. The gaugeon sector
  in (\ref{type2}) does not decouple for any value of $\alpha$.
\end{enumerate}
{}From the view point of 2, we may consider that Type I theory is preferable to
Type II theory. The property 1, however, implies that the framework of the
Type I theory is not wide enough, since we have to consider two different
Hilbert spaces corresponding to the sign of $a$.

The main purpose of the present paper is to extend the Type I gaugeon
formalism so that we can shift the gauge parameter $a$ freely into arbitrary
value. For this end, we introduce two sets of gaugeon fields $Y_i$ and
$Y_{i*}$ ($i=1,2$), each of which is accompanied by numerical parameter
$\alpha_i$.

The paper is organized as the following. In \S2, we propose an extended
gaugeon formalism which has the following properties: (1) the gauge
parameter $a$ appearing in the propagator can be shifted quite
freely by the q-number gauge transformation; (2) if we put
$\alpha_1=\alpha_2=0$,
the gaugeon sector of the Lagrangian decouples from the rest and the remaining
sector coincides with the Landau-gauge Lagrangian of the standard formalism;
(3) the theory is BRST symmetric.
In \S3, we see that our theory is a combined theory of Type I$_+$ and
Type I$_-$ gaugeon formalism. The relation to the Type II theory is discussed
in \S4.
By using various BRST charges we show that the Hilbert space of our
formalism has a subspace which can be identified with the
total space of Type II formalism. Section 5 is devoted to summary.

\section{Extended Type I theory}

We start from the Lagrangian given by
\begin{eqnarray}
    {\cal L} &=& {\cal L}_{\rm inv}
              + \partial_\mu B A^\mu
              + (Y_{1*} + \alpha_{1}B)(Y_{2*} + \alpha_{2}B)
              + \partial_\mu Y_{1*} \partial^\mu Y_{1}
              + \partial_\mu Y_{2*} \partial^\mu Y_{2}
                                                \nonumber \\
           & &- i \del_\mu c_* \del^\mu c
              - i \del_\mu K_{1*} \del^\mu K_{1}
              - i \del_\mu K_{2*} \del^\mu K_{2},
                                                 \label{L2Y}
\end{eqnarray}
where $c$ and $c_*$ are usual FP ghosts, $Y_i$ and $Y_{i*}$ ($i=1,2$) are
two sets of gaugeon fields, $K_i$ and $K_{i*}$ are FP ghosts for the gaugeon
fields \cite{Izawa,KSE} and $\alpha_i$ 's are the numerical parameters.
Field equations which follow from (\ref{L2Y}) are
\begin{eqnarray}
      &  &\del^\mu F_{\mu\nu} + \del_\nu B + j_\nu =0, \nonumber \\
      &  &\del^\mu A_\mu = 2\alpha_1\alpha_2 B
                           + \alpha_2 Y_{1*}
                           + \alpha_1 Y_{2*}, \nonumber \\
      &  &\Box Y_1 = Y_{2*} + \alpha_2 B, \nonumber\\
      &  &\Box Y_2 = Y_{1*} + \alpha_1 B, \nonumber\\
      &  &\Box Y_{i*} = 0, \nonumber\\
      &  &\Box c = \Box c_* = 0, \nonumber\\
      &  &\Box K_i = \Box K_{i*} = 0,
\end{eqnarray}
where $i=1,2$ and $j_\mu$ is the conserved current defined by
$j_\mu=\del{\cal L}_{\rm matt}/\del A^\mu$.

Our theory can be considered as an extension of Type I theory.
The Lagrangian (\ref{L2Y}) yields the photon propagator (\ref{propagator})
with
\begin{equation}
             a = 2 \alpha_1 \alpha_2.  \label{a2Y}
\end{equation}
Hence, if we put $\alpha_1=\varepsilon\alpha_2 (\equiv \alpha/\sqrt2)$, it
becomes $a=\varepsilon \alpha^2$, which is nothing but the expression for
$a$ (\ref{aI}) of Type I theory.
Furthermore, if we put $\alpha_1=\alpha_2=0$,
the gaugeon sector decouples from the rest in (\ref{L2Y}), and the remaining
sector is identical with the Landau-gauge Lagrangian of the standard
formalism. This is the characteristic of the Type I theory.

The Lagrangian (\ref{L2Y}) is invariant under the following BRST
transformation:
\begin{eqnarray}
  & &\delta_{\rm B} A_\mu = \partial_\mu c, \nonumber \\
  & &\delta_{\rm B} \psi = iec\psi,         \nonumber \\
  & &\delta_{\rm B} c_* = -iB,              \nonumber \\
  & &\delta_{\rm B} B = \delta_{\rm B}c=0,  \nonumber \\
  & &\delta_{\rm  B} Y_i = K_{i},           \nonumber \\
  & &\delta_{\rm B} K_{i*} = -iY_{i*},      \nonumber \\
  & &\delta_{\rm B}Y_{i*} =\delta_{\rm B}K_{i}=0 \quad (i=1, 2)
                                            \label{BRST}
\end{eqnarray}
which obviously satisfies the nilpotency, $\delta_{\rm B}^{~2}=0$.
Because of the nilpotency, the BRST invariance is
easily seen if we rewrite the Lagrangian as
\begin{eqnarray}
  {\cal L} &=& {\cal L}_{\rm inv} + i \delta_{\rm B}
         \left[
                 \partial_\mu c_* A^\mu
               + \partial_\mu K_{1*} \partial^\mu Y_1
               + \partial_\mu K_{2*} \partial^\mu Y_2 {{}\over{}} \right.
                                                         \nonumber \\
     & &\left. + {1\over2}(K_{1*} + \alpha_1 c_*)( Y_{2*} + \alpha_2 B)
               - {1\over2}(K_{2*} + \alpha_2 c_*)( Y_{1*} + \alpha_1 B)
         \right].
\end{eqnarray}
BRST charge corresponding to this transformation is expressed by
\begin{equation}
    Q_{\rm B} = \int ( c \mathop{\del_0}^\leftrightarrow B
                           + K_1 \mathop{\del_0}^\leftrightarrow Y_{1*}
                           + K_2 \mathop{\del_0}^\leftrightarrow Y_{2*}
                                   )d^3x,             \label{QB}
\end{equation}
with
$      \stackrel{\leftrightarrow}{\del_0}
      = \del_0 - \stackrel{\leftarrow}{\del_0}$.
By the help of this charge we can define the physical subspace
${\cal V}_{\rm phys}$ as a space of states satisfying
\begin{equation}
     Q_{\rm B} \left|{\rm phys} \right\rangle = 0.  \label{QBphys}
\end{equation}
This subsidiary condition removes the gaugeon modes as well as the
unphysical photons from the physical subspace; $Y_i$ and $Y_{i*}$
together with $K_i$ and $K_{i*}$ constitute BRST quartets.

We define the q-number gauge transformation by
\begin{eqnarray}
   & &\hat A_\mu = A_\mu
                   + \tau_1 \partial_\mu Y_1
                   + \tau_2 \partial_\mu Y_2,            \nonumber \\
   & &\hat \psi = \exp[ie(\tau_1 Y_1 +\tau_2 Y_2)]\psi,  \nonumber \\
   & &\hat Y_{i*} = Y_{i*} - \tau_i B,                   \nonumber \\
   & &\hat B = B, \quad \quad \hat Y_i = Y_i,            \nonumber \\
   & &\hat c = c + \tau_1 K_1 + \tau_2 K_2,              \nonumber \\
   & &\hat K_{i*} = K_{i*} - \tau_i c_*,  	         \nonumber \\
   & &\hat c_* = c_*, \quad \quad \hat K_i = K_i, \quad \quad (i=1, 2)
                                                         \label{qgt2Y}
\end{eqnarray}
where $\tau_i$'s are the parameters of the transformation.
Under this transformation, the Lagrangian is form invariant:
\begin{equation}
             {\cal L}(\phi^A; \alpha_1, \alpha_2)
                =  {\cal L}(\hat \phi^A; \hat \alpha_1, \hat \alpha_2),
                                \label{forminv2Y}
\end{equation}
where $\phi^A$ stands for any of the fields and $\hat \alpha_i$'s are
defined by
\begin{equation}
      \hat \alpha_i = \alpha_i + \tau_i. \quad (i=1,2)
\end{equation}

Note that this q-number gauge transformation commutes with the BRST
transformation (\ref{BRST}). As a result, the BRST charge (\ref{QB}) is
invariant under the q-number transformation,
\begin{equation}
        \hat Q_{\rm B} = Q_{\rm B},  \label{QB_hat}
\end{equation}
and therefore the physical subspace ${\cal V}_{\rm phys}$ is also invariant:
\begin{equation}
             \hat {\cal V}_{\rm phys} = {\cal V}_{\rm phys}.
\end{equation}

\section{Relation to Type I theory}

As mentioned in last section, our theory can be considered as an extension of
Type I gaugeon formalism. This will be understood more clearly if we rewrite
the Lagrangian as the following. Let us consider the field redefinition as
\begin{eqnarray}
       & &Y_{\pm}  = {1\over\sqrt2}(Y_{1} \pm Y_{2} ), \nonumber \\
       & &Y_{\pm*}  = {1\over\sqrt2}(Y_{1*} \pm Y_{2*} ),
\end{eqnarray}
and by similar definition we introduce $K_{\pm}$ and $K_{\pm*}$.
Using these fields the Lagrangian (\ref{L2Y}) can be written as
\begin{eqnarray}
    {\cal L} &=& {\cal L}_{\rm inv}
              + \partial_\mu B A^\mu
              + {1 \over 2} \left( Y_{+*} + \alpha_+ B \right)^2
              - {1 \over 2} \left( Y_{-*} + \alpha_- B \right)^2
              + \partial_\mu Y_{+*} \partial^\mu Y_+
                                            \nonumber \\
           & &+ \partial_\mu Y_{-*} \partial^\mu Y_-
              - i \del_\mu c_* \del^\mu c
              - i \del_\mu K_{+*} \del^\mu K_+
              - i \del_\mu K_{-*} \del^\mu K_-
                                                 \label{L2Y+-}
\end{eqnarray}
with parameters $\alpha_{\pm}$ defined by
\begin{equation}
         \alpha_{\pm} = {1\over \sqrt2}(\alpha_1 \pm \alpha_2).
\end{equation}
The q-number gauge transformation (\ref{qgt2Y}) becomes
\begin{eqnarray}
   & &\hat A_\mu = A_\mu
                   + \tau_+ \partial_\mu Y_+
                   + \tau_- \partial_\mu Y_-,            \nonumber \\
   & &\hat \psi = \exp[ie(\tau_+ Y_+ +\tau_- Y_-)]\psi,  \nonumber \\
   & &\hat Y_{\pm*} = Y_{\pm*} - \tau_{\pm} B,           \nonumber \\
   & &\hat B = B, \quad \quad \hat Y_{\pm} = Y_{\pm},    \nonumber \\
   & &\hat c = c + \tau_+ K_+ + \tau_- K_-,              \nonumber \\
   & &\hat K_{\pm*} = K_{\pm*} - \tau_{\pm} c_*,  	 \nonumber \\
   & &\hat c_* = c_*, \quad \quad \hat K_{\pm} = K_{\pm},
                                                         \label{qgt2Y+-}
\end{eqnarray}
together with the parameter shifts
\begin{equation}
       \hat \alpha_\pm = \alpha_\pm + \tau_\pm,
\end{equation}
where the parameters of the transformation $\tau_\pm$ are given by
$\tau_\pm=(\tau_1 \pm \tau_2)/\sqrt2$.
As seen from (\ref{L2Y+-}) and (\ref{qgt2Y+-}), $Y_+$, $Y_{+*}$, $K_+$ and
$K_{+*}$ [$Y_-$, $Y_{-*}$, $K_-$ and $K_{-*}$] can be identified with the
gaugeon fields
and their FP ghosts of Type I$_+$ [Type I$_-$] formalism. Namely, our theory
is a combined theory of Type I$_+$ and Type I$_-$ gaugeon formalism.
Especially, the gauge parameter $a$ appearing in the propagator is given by
\begin{equation}
     a = \alpha_+^{~2} - \alpha_-^{~2},
\end{equation}
which can take both signs.

\section{Relation to Type II theory}

We have seen in the preceding sections that in our extended Type I theory
the gauge parameter $a$ appearing in the propagator can take both signs.
This is the characteristic of Type II theory. So, one may expect that the Type
II theory is included as a kind of sub-theory in the present formalism. We
discuss below such relation.

Remembering that the gaugeon sector never decouples in Type II formalism, we
assume that at least one of the parameters $\alpha_1$ and $\alpha_2$ in
(\ref{L2Y}) differs from zero. For simplicity, we assume $\alpha_1\neq 0$
in this section. Then, we can consider the following field rescaling:
\begin{eqnarray}
     & &\phi^A \rightarrow (2\alpha_1)^{-1} \phi^A, \quad
         (\mbox{for } \phi^A=Y_1, K_1, Y_{2*}, K_{2*})
                                                     \nonumber \\
     & &\phi^A \rightarrow (2\alpha_1) \phi^A. \quad
         (\mbox{for } \phi^A=Y_{1*}, K_{1*}, Y_{2}, K_{2})
                                                   \label{rescale}
\end{eqnarray}
Under this rescaling, $Y_{1*}Y_{2*}$-term and kinetic terms for gaugeon fields
are invariant and the Lagrangian becomes
\begin{eqnarray}
    {\cal L} &=&{\cal L}_{\rm inv}
              + \partial_\mu B A^\mu
              + {\alpha \over 2}B^2
              + \alpha B Y_{1*}
              + {1\over2} B Y_{2*}
              + Y_{1*} Y_{2*}
                                            \nonumber \\
           & &+ \partial_\mu Y_{1*} \partial^\mu Y_1
              + \partial_\mu Y_{2*} \partial^\mu Y_2
              - i \del_\mu c_* \del^\mu c
              - i \del_\mu K_{1*} \del^\mu K_1
              - i \del_\mu K_{2*} \del^\mu K_2.
                                                 \label{L2Y2}
\end{eqnarray}
Here we have introduced $\alpha$ defined by
\begin{equation}
      \alpha = 2 \alpha_1 \alpha_2 = \alpha_{+}^{~2} - \alpha_{-}^{~2},
\end{equation}
since the parameters $\alpha_1$ and $\alpha_2$ appear only in this combination
after the field rescaling.
Note that one may obtain the Lagrangian (\ref{L2Y2}) also by taking special
values for $\alpha_{1,2}$ in (\ref{L2Y}) as
\begin{equation}
     \alpha_1=1/2~(\neq0), \quad \quad \alpha_2=\alpha.  \label{alpha1=1/2}
\end{equation}

Now let us consider the q-number gauge transformation. Since we have assumed
$\alpha_1 \neq 0$ (or $\alpha_1=1/2$), we restrict the transformation that
does not shift $\alpha_1$. Then, the transformation is given by
\begin{eqnarray}
   & &\hat A_\mu = A_\mu
                   + \tau \partial_\mu Y_2,   \nonumber \\
   & &\hat \psi = \exp[ie\tau Y_2]\psi,       \nonumber \\
   & &\hat Y_{2*} = Y_{2*} - \tau B,          \nonumber \\
   & &\hat B = B, \quad \hat Y_2 = Y_2,       \nonumber \\
   & &\hat c = c + \tau K_2,                  \nonumber \\
   & &\hat K_{2*} = K_{2*} - \tau c_*,        \nonumber \\
   & &\hat c_* = c_*, \quad \hat K_2 = K_2,   \nonumber \\
   & &\hat Y_1=Y_1, \quad \hat Y_{1*} = Y_{1*}, \nonumber \\
   & &\hat K_1=K_1, \quad \hat K_{1*} = K_{1*},
                                                         \label{qgt2Y2}
\end{eqnarray}
under which the parameter $\alpha$ shifts as
\begin{equation}
           \hat \alpha = \alpha + \tau,
\end{equation}
where the parameter $\tau$ of the transformation is related to $\tau_2$ by
$\tau=2\alpha_1 \tau_2$.

Since the field rescaling (\ref{rescale}) commutes with the BRST
transformation, the Lagrangian (\ref{L2Y2}) is invariant under (\ref{BRST}).
We decompose this BRST transformation as
$$
     \delta_{\rm B} = \delta_{\rm B(KO)}
                    + \delta_{\rm B(Y1)}
                    + \delta_{\rm B(Y2)},
$$
where $\delta_{\rm B(KO)}$ denotes $\delta_{\rm B}$ that acts only on $A_\mu$,
$\psi$, $B$, $c$, and $c_*$ fields, and $\delta_{{\rm B(Y}i)}$ acts only on
$Y_i$, $Y_{i*}$, $K_i$ and $K_{i*}$ ($i=1,2$). The Lagrangian is invariant
separately under these BRST transformations. The corresponding
conserved BRST charges are given by
\begin{eqnarray}
  & &Q_{\rm B(KO)} = \int c \stackrel{\leftrightarrow}{\del_0} B d^3x,
				      \nonumber \\
  & &Q_{\rm B(Y1)} = \int K_1 \stackrel{\leftrightarrow}{\del_0} Y_{1*} d^3x,
      					\nonumber \\
  & &Q_{\rm B(Y2)} = \int K_2 \stackrel{\leftrightarrow}{\del_0} Y_{2*} d^3x.
\end{eqnarray}
In \S2, we have taken (\ref{QBphys}) as a physical subsidiary
condition. Instead of it, we may choose the condition as
\begin{eqnarray}
   & &(Q_{\rm B(KO)} + Q_{\rm B(Y2)})
         \left| {\rm phys} \right\rangle = 0, \nonumber \\
   & &Q_{\rm B(Y1)} \left| {\rm phys} \right\rangle = 0. \label{QBY12phys}
\end{eqnarray}
We denote the space of states satisfying (\ref{QBY12phys}) by
${\cal V}_{\rm phys}^{\rm (II)}$. It can be easily seen that this space
is a subspace of ${\cal V}_{\rm phys}$ defined in \S2:
\begin{equation}
          {\cal V}_{\rm phys}^{\rm (II)}
                        \subset {\cal V}_{\rm phys}. \label{VsubV}
\end{equation}
As seen below, the space ${\cal V}_{\rm phys}^{\rm (II)}$ can be considered
as a physical subspace of BRST symmetric Type II theory \cite{Izawa}.
Furthermore, if we define a subspace ${\cal V}^{\rm (II)}$ of total space
${\cal V}$ by
\begin{equation}
   {\cal V}^{\rm (II)} = \{ \left| \Phi \right\rangle\in {\cal V} ; ~
                Q_{\rm B(Y1)} \left| \Phi \right\rangle =0 \}
                \subset {\cal V},
\end{equation}
then this space ${\cal V}^{\rm (II)}$ can be identified with the total space
of Type II formalism. In particular, the physical subspace
${\cal V}_{\rm phys}^{\rm (II)}$ cab be expressed by
\begin{equation}
   {\cal V}_{\rm phys}^{\rm (II)}=
   \{ \left| \Phi \right\rangle\in {\cal V}^{\rm (II)} ; ~
                (Q_{\rm B(KO)} + Q_{\rm B(Y2)})
                \left| \Phi \right\rangle =0 \}
                \subset {\cal V}^{\rm (II)}.
                                               \label{QBIIphys}
\end{equation}

Let us show the statement mentioned above. We notice the following three
facts:
\begin{enumerate}
  \item
   The equal-time commutation relations for the fields $A_\mu$, $B$, $c$,
   $c_*$, $\psi$, $Y_2$, $Y_{2*}$, $K_2$ and $K_{2*}$ are exactly the same
   with those of the Type II formalism.
  \item
   If we take the matrix elements among the states of ${\cal V}^{\rm (II)}$,
   $A_\mu$, $B$, $c$, $c_*$, $\psi$, $Y_2$, $Y_{2*}$, $K_2$ and $K_{2*}$
   satisfy the same field equations with those for the Type II formalism.
  \item
   Any state given by a product of the field operators $A_\mu$, $B$, $c$,
   $c_*$, $\psi$, $Y_2$, $Y_{2*}$, $K_2$ and $K_{2*}$ acting on the vacuum
   state is included in ${\cal V}^{\rm (II)}$, since these fields are
   $Q_{\rm B(Y1)}$-singlets.
\end{enumerate}
The second fact can be easily understood if we express the Lagrangian
(\ref{L2Y2}) as
\begin{eqnarray}
  {\cal L} &=& {\cal L}_{\rm inv}
               + \partial^\mu A_\mu
               + {\alpha \over 2} B^2
               + {1 \over 2} Y_{2*} B
               + \partial_\mu Y_{2*} \partial^\mu Y_2
                                                \nonumber \\
           & & + i \delta_{\rm B(Y1)}
         \left[
                 \partial_\mu K_{1*} \partial^\mu Y_1
               + K_{1*}( Y_{2*} + \alpha B)
         \right],
                                            \label{L2YQB1}
\end{eqnarray}
which is equivalent to the Type II Lagrangian up to null operator over
${\cal V}^{\rm (II)}$.
The facts 1 and 2 mean that the field equations and the (four-dimensional)
commutation relations are the same with those of the Type II formalism
if their matrix elements are assumed to be taken
in ${\cal V}^{\rm (II)}$. Combining this with 3, we can conclude that
any vacuum expectation value of the products of $A_\mu$, $B$, $c$, $c_*$,
$\psi$, $Y_2$, $Y_{2*}$, $K_2$ and $K_{2*}$ fields coincides with that
evaluated in the Type II formalism.

\section{Summary}

By introducing two kinds of gaugeon fields $Y_i$ and $Y_{i*}$ and two
numerical parameters $\alpha_i$ ($i=1,2$), we have proposed an extended
Type I gaugeon formalism for quantum electrodynamics.
The gauge parameter $a=\varepsilon\alpha^2$ appearing in the propagator
has been replaced by $a=2\alpha_1\alpha_2$ through this extension.
As a result, the parameter $a$ can be shifted into arbitrary value by the
q-number gauge transformation in the present formalism.

{}From the original Type I gaugeon formalism, our theory inherits the
following property: If we put $\alpha_1=\alpha_2=0$, the gaugeon sector in
the Lagrangian decouples from the rest and the theory becomes manifestly
equivalent to the Landau-gauge standard formalism. This property
can be easily understood since, as shown in \S3, our theory is
a combined theory of Type I$_+$ and Type I$_-$ gaugeon formalism.

The present formalism also includes Type II theory in the following sense.
By the help of BRST-like symmetries,
we have defined a subspace ${\cal V}^{\rm (II)}$ which can be identified with
the total space of the Type II theory. Thus the Hilbert space of the Type II
theory is embedded in that of the present formalism.


\end{document}